\documentclass[pss]{wiley2sp} 
\usepackage{amsmath}

\begin{document}

\title{Superconductivity-enhanced Nematicity and '$s+d$' Gap Symmetry in Fe(Se$_{1-x}$S$_x$)}

\titlerunning{Superconductivity-enhanced Nematicity and '$s+d$' Gap Symmetry in Fe(Se$_{1-x}$S$_x$)}

\author{%
  Liran Wang\textsuperscript{\Ast,\textsf{\bfseries 1}},
  Fr\'ed\'eric Hardy\textsuperscript{\textsf{\bfseries 1}},
  Thomas Wolf\textsuperscript{\textsf{\bfseries 1}},
  Peter Adelmann\textsuperscript{\textsf{\bfseries 1}}, 
  Rainer Fromknecht\textsuperscript{\textsf{\bfseries 1}},  
   Peter Schweiss\textsuperscript{\textsf{\bfseries 1}}, 
   Christoph Meingast\textsuperscript{\textsf{\bfseries 1}}, 
}

\authorrunning{Liran Wang et al.}

\mail{e-mail
  \textsf{liran.wang@kit.edu}, Phone:
  +49 (0)721608-24719, Fax: +49 (0)721608-24624}

\institute{%
  \textsuperscript{1}\,Institute for Solid-State Physics, Karlsruhe Institute
of Technology, 76021 Karlsruhe, Germany}

\received{XXXX, revised XXXX, accepted XXXX} 
\published{XXXX} 

\keywords{Thermal expansion, heat capacity, S-substituted FeSe, nematicity, supercoductivity}

\abstract{%
%
%
%
\abstcol{%
Superconducting iron chalcogenide FeSe has the simplest crystal structure among all the Fe-based superconductors. Unlike other iron pnictides, FeSe exhibits no long range magnetic order accompanying the tetragonal-to-orthorhombic structural distortion, which raises the fundamental question about the role of magnetism and its associated spin fluctuations in mediating both nematicity and superconductivity. The extreme sensitivity of FeSe to external pressure suggests that chemical pressure, induced by substitution of Se by the smaller ion S, could also a be good tuning parameter to further study the coupling between superconductivity and nematicity and to obtain information on both the Fermi-surface changes and the symmetry of the superconducting state.} {Here we study the thermodynamic properties of Fe(Se$_{1-x}$S$_{x}$) for 3 compositions, $x=0$, 0.08 and 0.15, using heat-capacity and thermal-expansion measurements. With increasing S content we observe a significant reduction of the tetragonal-to-orthorhombic transition temperature T$_{s}$. However, this suppression of T$_{s}$ is counterintuitively accompanied by an enhancement of the orthorhombic distortion $\delta$ below T$_c$, which clearly indicates that superconductivity favors the nematic state. In parallel, the superconducting transition temperature T$_{c}$ is sizeably enhanced, whereas the increase of the Sommerfeld coefficient $\gamma_{n}$ is quite moderate. In the T$\to$ 0 limit, an unusually large residual density of states is found for $x>0$ indicative of significant substitution-induced disorder. We discuss these observations in the context of $s+d$ superconducting-state symmetry.}}

%
%


\maketitle   

\section{Introduction}
Recently, iron chalcogenide FeSe (11 system) has attracted growing attention because of its unusual features. Indeed, the long range magnetic order, which is closely linked to the tetragonal-to-orthorhombic transition T$_{s}$ in 1111 ({\it e.g.} : LaFeAsO) \cite{McGuirePRBsRevB2008,LiranPRB2009} and 122 ({\it e.g.} : BaFe$_{2}$As$_{2}$) \cite{Johnston:2010aa,FredEurophyL2010} iron-pnictide systems, is absent in FeSe at ambient pressure \cite{ImaiPRL2009}. This distorted non-magnetic phase has been coined 'nematic' and whether the origin is due to orbital or magnetic degrees of freedom is currently under strong debate \cite{Fernandes2014,Bohmer:2015ab,AnnaPRB2013,Baek:2015aa}. Under hydrostatic pressure, T$_{s}$ is progressively suppressed and a new high-pressure phase, likely magnetic, emerges, while T$_{c}$ is strongly enhanced from 8.5 K at P =0 to about 37 K at P $\approx$ 9 GPa \cite{Medvedev2009,PhysRevB.80.064506,UdharaPRB2016,TerashimaPRB2016,TerashimaJPSJ2015,2015arXiv151206951S,KnonerPRB2015}. Surprisingly, T$_{c}$ can be further increased to 50 K in monolayer FeSe  deposited on SrTiO$_3$ substrates \cite{JFGeNM2015}. This extreme sensitivity of FeSe to external pressure leads to a complex (T, P) phase diagram \cite{TerashimaPRB2016,TerashimaJPSJ2015,2015arXiv151206951S}, which is somewhat reminiscent of the complicated (T, $x$) phase diagrams of hole doped Ba122 systems \cite{LiranWangPRB2016,Bohmer:2015ab}. This suggests that chemical pressure, induced by substitution of Se by the smaller ion S, could also be a good tuning parameter to further study the coupling between superconductivity and nematicity and to obtain information on both the Fermi-surface changes and the symmetry of the superconducting state \cite{MizuguchiJPSJ2009,WatsonPRB2015,MahmoudPRB2015,HosoiarXiv2016}.\\

In this paper, we investigate the thermodynamic properties of sulfur-substituted FeSe single crystals, Fe(Se$_{1-x}$S$_{x}$), for 3 compositions $x=$ 0, 0.08 and $\sim$0.15, using heat-capacity and thermal-expansion measurements. In agreement with previous studies \cite{WatsonPRB2015,HosoiarXiv2016}, we find that T$_{s}$ is strongly suppressed with increasing S content. By studying the temperature dependence of the orthorhombic distortion, we surprisingly find that, in contrast to most other Fe-based systems, superconductivity favors the distorted nematic state.  This is in qualitative agreement to recent theoretical work \cite{KangPRL2014}, which predicts that orbital order increases T$_c$, but is in strong contrast to the behavior found in other Fe-based systems. The heat capacity data show that S substitution leads to a slight increase in the electronic density of states and results in a significant residual density of states at T = 0 K.

%

\section{Experimental methods}
\paragraph{Crystal growth and Characterization}
Single crystals of FeSe$_{1-x}$S$_{x}$ were grown in evacuated quartz ampoules by the KCl/AlCl$_{3}$ chemical vapor transport method \cite{AnnaPRB2013,WatsonPRB2015}. Single crystals with a mass of 1-3 mg were chosen for thermal-expansion and specific-heat measurements. The composition of two of the samples was determined by refinement of four-circle single-crystal X-ray diffraction patterns of a small piece of each crystal to be $x$ = 0 and 0.08.  The composition of the third sample was estimated to be $\sim$0.15 from the value of T$_s$\cite{WatsonPRB2015}.  
\paragraph{Thermodynamic measurements}\label{thermo}
Thermal expansion was measured using a high-resolution home-made capacitance dilatometer\cite{Meingast1990}, which is several orders of magnitude more sensitive than traditional diffraction techniques.The thermal expansion, $\Delta L/L$, of the individual $a$ and $b$ axes, as well as the distortion $\delta=|a-b|/(a+b)$, can be derived using the difference between 'twinned' and 'detwinned' datasets. Indeed, the shorter $b$-axis in the low-temperature orthorhombic phase can be obtained directly by measuring the expansion of the crystal along the {[}110{]}\textsubscript{T} direction of the original tetragonal cell, because in this configuration the small force from the dilatometer detwins the crystal (see the insert of Fig.\ref{Fig1}a)\cite{Bohmer:2015ab,LiranWangPRB2016,Boehmer2012PRB}. The larger $a$-axis, on the other hand, is obtained by combining a 'twinned' measurement, along {[}100{]}\textsubscript{T}, with the 'detwinned' data.
Heat capacity was measured using a Physical Property Measurement System (PPMS) from Quantum Design. Data in constant magnetic field (H = 0 and 14 T) were obtained using the dual-slope method \cite{CMarcenatPhDthesis,Riegel1986}.

\section{Results and discussion}

\begin{figure*}[htb]%
\includegraphics*[width=\textwidth]{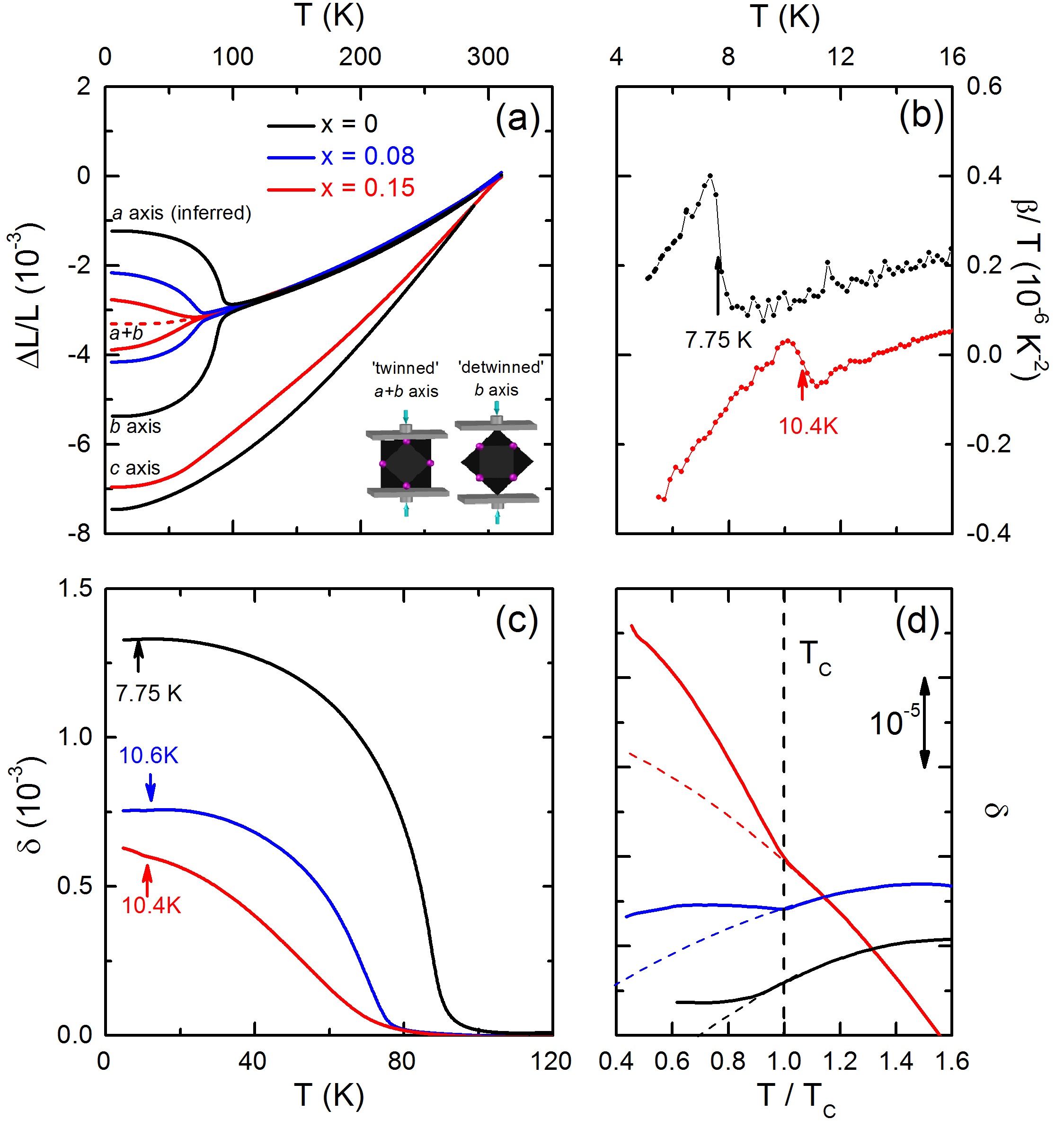}
\caption{(a) Temperature dependence of the relative length change $\Delta L/L$ of the orthorhombic lattice parameters $a$, $b$ and $c$ of FeSe$_{1-x}$S$_{x}$ for $x$ = 0, 0.08, 0.15. The dashed line shows the 'twinned' measurement for $x$ = 0.15. The inset shows the two configurations  {[}110{]}\textsubscript{T} and  {[}100{]}\textsubscript{T} (see Section \ref{thermo} for details). (b) Temperature dependence of the coefficient of volume thermal expansion for $x$ = 0.0 and $x$ = 0.15. The arrows indicate the superconducting transition. (c) Temperature dependence of the orthorhombic distortion, $\delta=|a-b|/(a+b)$, inferred from the data in (a). (d) Enlarged view of the low-temperature distortion in (c). Data are shifted with respect to each other. The dashed lines represents extrapolations of the normal-state data below T$_c$.}
\label{Fig1}
\end{figure*}


\paragraph{Interplay between nematicity (orthorhombicity) and superconductivity}
Figure \ref{Fig1}a shows the relative length changes $\Delta L/L$ along the $a$ and $b$ axes for the three FeSe$_{1-x}$S$_x$ single crystals with composition $x$ = 0, 0.08, 0.15. Results for the $c$ axis are also given for
$x$ = 0 and 0.15. The expected orthorhombic splitting of the $a$ and $b$ lattice parameters is clearly observed below T$_{s}$.  We find that both T$_{s}$ and the splitting are reduced in magnitude with increasing S content. The coupling of superconductivity to the lattice parameters is not obvious in Fig. \ref{Fig1}a due to the smallness of these effects. However, a sizeable jump is clearly resolved at T$_{c}$ in the volume  thermal expansion coefficient $\beta$(T) = 1/VdV/dT, as illustrated in Fig.\ref{Fig1}b, implying well-defined bulk superconductivity for S-substituted FeSe. Here, $\beta$(T) is just the sum of the individual linear expansion coefficients. In contrast to T$_{s}$, T$_{c}$ increases moderately with S substitution, as has been previously observed \cite{MizuguchiJPSJ2009,WatsonPRB2015,MahmoudPRB2015}.

In Fig. \ref{Fig1}c, we show the orthorhombic distortion $\delta$=$\vert a-b \vert$/(b+a) calculated from the thermal-expansion data from Fig. 1a. The orthorhombic splitting is reduced by a factor of $\approx$ 3 between $x$ = 0 and $x$ = 0.15. Recent ARPES measurements \cite{WatsonPRB2015} indicate that this suppression of orthorhombicity is strongly correlated to the reduction of the $d_{xz}/d_{yz}$ splitting, suggesting that orbital ordering is likely the driving force for the structural transition in the absence of magnetism. The effect of superconductivity on $\delta$ can be more clearly seen in Fig. \ref{Fig1}d, which is a magnification of the low-temperature region. Here, a kink in the data indicates the onset of superconductivity. We note that there is no kink at T$_c$ in pure FeSe \cite{AnnaPRB2013}. Remarkably, we find that the distortion for T$<$ T$_{c}$ is enhanced from the value expected by extrapolation from the normal state (dashed line) and that this effect becomes even more prominent for larger S contents. This result is totally opposite to what has been observed in underdoped Ba(Fe$_{1-x}$Co$_{x}$)$_{2}$As$_{2}$ \cite{NandiPRL2010,ChristophPRL2012}, Ba$_{1-x}$K$_{x}$Fe$_{2}$As$_{2}$\cite{Bohmer:2015ab} and BaFe$_{2}$(As$_{1-x}$P)$_{x}$ \cite{Boehmer2012PRB} where $\delta$ is reduced below T$_{c}$, which has been interpreted as a sign of strong competition between magnetism and superconductivity.  Though some spin fluctuation remains in FeSe at low T \cite{Boehmer:2015aa,2015arXiv151104757T},  this enhancement confirms that superconductivity and nematicity (orthorhombicity) do not compete in FeSe$_{1-x}$S$_x$. On the contrary, it appears that superconductivity favors the lower symmetry structure.

\begin{figure*}[htb]%
\includegraphics*[width=\textwidth]{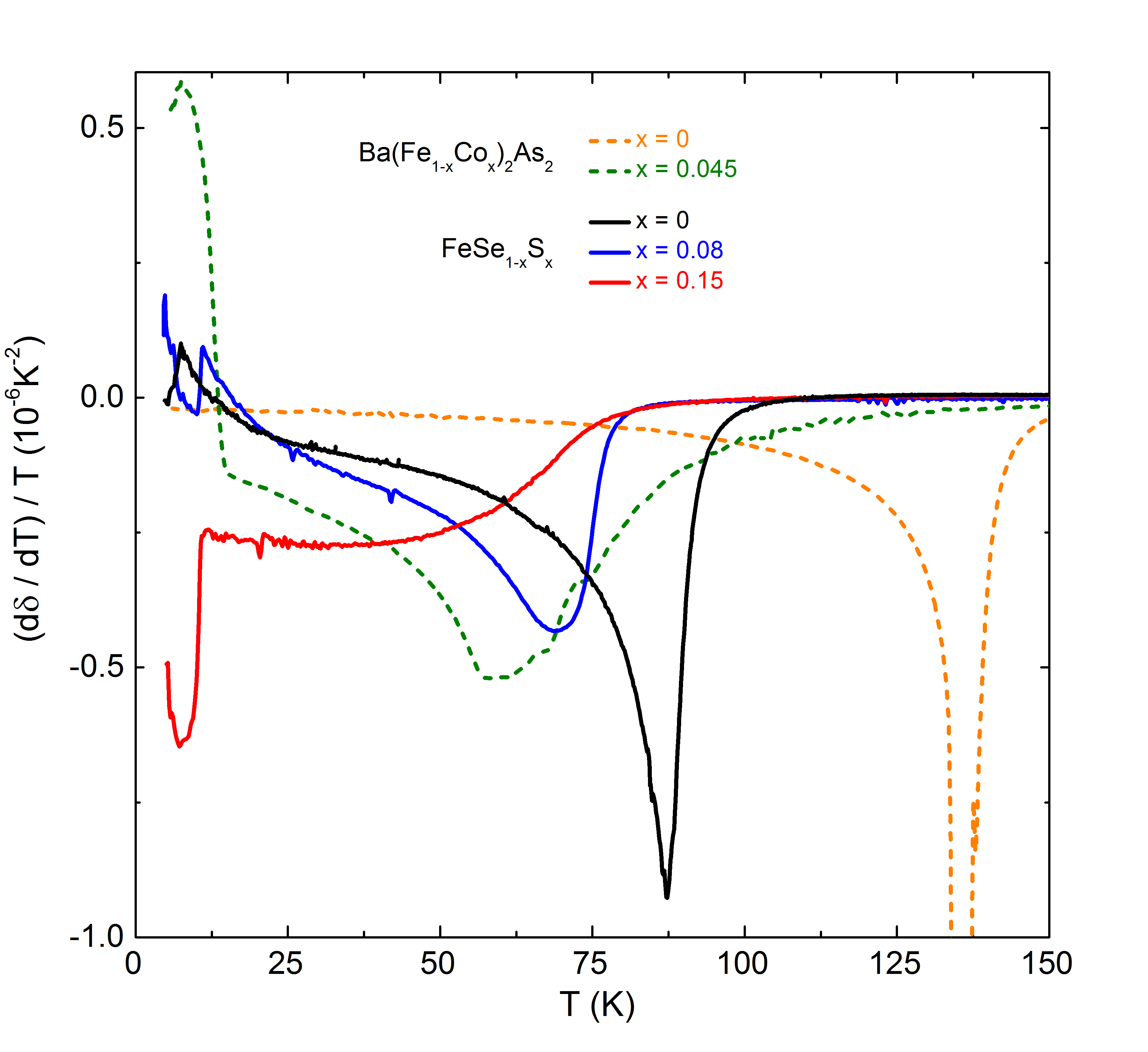}
\caption{Temperature dependence of $(d\delta/dT)/T$ for FeSe$_{1-x}$S$_{x}$ ($x$ = 0, 0.08, 0.15) and Ba(Fe$_{1-x}$Co$_x$)$_2$As$_2$ ($x$ = 0, 0.045).}
\label{Fig2}
\end{figure*}

The above effects are more clearly seen in Fig.\ref{Fig2}, where we compare the temperature derivative of $\delta$/T of FeSe$_{1-x}$S$_x$ (solid lines) to that of Ba(Fe$_{1-x}$Co$_x$)$_2$As$_2$ ($x$ = 0, 0.045) (dashed lines). Similar to FeSe$_{1-x}$S$_x$, T$_{s}$ is  suppressed with increasing Co substitution in BaFe$_{2}$As$_2$. However, the behavior for T $<$ T$_{c}$ is fundamentally different. Indeed, the competition between superconductivity and orthorhombicity in Ba(Fe$_{1-x}$Co$_x$)$_2$As$_2$ is exemplified by a reduction of $\delta$ below T$_{c}$, or equivalently by a positive anomaly in $d\delta/dT$. The distortion in Ba(Fe$_{1-x}$Co$_x$)$_2$As$_2$  is induced by the magnetic order via a large magneto-elastic coupling, and the subsequent reduction of $\delta$ below T$_{c}$ indicates the weakening of magnetism by superconductivity \cite{NandiPRL2010,ChristophPRL2012}. As shown in Fig.\ref{Fig2}, the slope change of $\delta$ at T$_{c}$ in FeSe$_{1-x}$S$_x$ is opposite in sign and indicates a rather different coupling mechanism in the absence of magnetism.  Interestingly, our result is in qualitative agreement to recent theoretical work \cite{KangPRL2014}, which predicts that T$_c$ will be enhanced in a distorted, orbitally-ordered non-magnetic state.

\begin{figure*}[htb]%
  \sidecaption
  \includegraphics*[width=\textwidth,height=15cm]{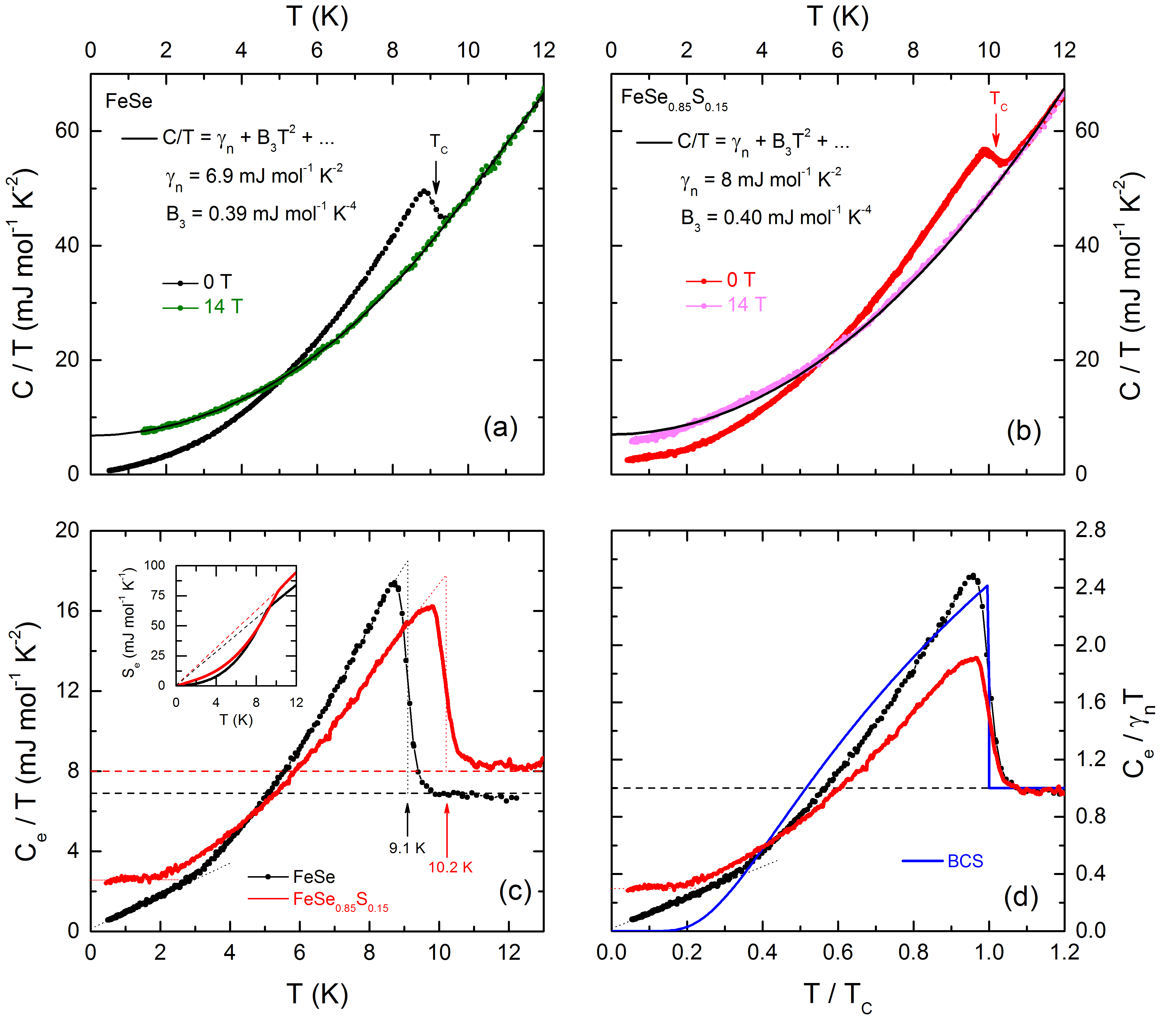}%
  \caption[]{%
Low temperature specific-heat of (a) FeSe and (b) Fe(Se$_{0.85}$S$_{0.15}$) in 0 and 14T, respectively. The black solid lines represent  fits to the H=14T data, which are used to extract the normal-state electronic and lattice contributions (see text). (c)-(d) Electronic specific heat for both compositions. The dashed line indicates a conserving-entropy construction to estimate T$_c$. The inset in c) shows the electronic entropy. The blue solid line in d) shows the BCS curve for a single-gap $s$-wave superconductor.}
    \label{Fig3}
\end{figure*}

\paragraph{Heat Capacity in the Superconducting State}
Figures \ref{Fig3}a and \ref{Fig3}b show the low-temperature specific heat of  single crystals with composition $x$ = 0 and 0.15, respectively, in H = 0 and 14 T applied along the $c$ direction. For both samples a clear anomaly indicates the transition to the superconducting state in the raw data.  To extract the Sommerfeld coefficient $\gamma_{n}$, the H = 14 T data were fitted to
\begin{equation}
\label{eq1}
C(T)=\gamma_nT+B_3T^3+B_5T^5+B_7T^7,
\end{equation}
where $\gamma_{n}T$ is the normal-state electronic contribution and the polynomials describes the lattice heat capacity within the harmonic-lattice approximation \cite{NormanCRC1971} (see black line in Figs \ref{Fig3}a and \ref{Fig3}b). Superconductivity is fully  suppressed in H = 14 T for $x$ = 0, while traces are still found below 3 K for $x$ = 0.15, as indicated by the deviations of the data to the Debye fit (see Fig. \ref{Fig3}b).We find that $\gamma_n$ slightly increases with S substitution from 6.9 mJ mol$^{-1}$ K$^{-2}$ for $x$ = 0 to $\approx$ 8 mJ mol$^{-1}$ K$^{-2}$ for $x$ = 0.15 indicating only minor changes of the Fermi surface as observed by recent photoemission spectroscopy \cite{WatsonPRB2015}. A similar Debye term B$_{3} \approx 0.4$ mJ mol$^{-1}$ K$^{-4}$ is inferred for both compounds leading to a Debye temperature $\theta_{D}$ $\approx$ 170 K. The resulting electronic heat capacity C$_{e}$(T) down to 0.4 K is illustrated in Figs \ref{Fig3}c and \ref{Fig3}d, and the inset shows that the thermodynamic requirement, that the superconducting-state entropy equals the normal-state entropy at T = T$_{c}$, is fulfilled. We find that the moderate enhancement of $\gamma_{n}$ is accompanied by a small increase of T$_{c}$ by about 1 K. Conversely, the heat-capacity jump $\Delta C / \gamma_{n}T_{c}$, which amounts to 1.68 in FeSe, indicating some strong-coupling effects, decreases to 1.24 for $x$ = 0.15. As discussed later, this is ascribed to an increasing amount of disorder induced by the S substitution.

Figure \ref{Fig3}d demonstrates that C$_{e}$(T) in FeSe strongly deviates from the behavior expected for a single-band $s$-wave BCS superconductor. In particular, C$_{e}$/T decreases nearly linearly for T / T$_{c}$ $<$ 0.3 and extrapolates, at T = 0, to a negligibly small residual value (see dotted black line). These observations are consistent with the existence of low-energy quasi-particle excitations related either to line nodes in $\Delta$({\bf k}) or to deep gap minima, at least, on one Fermi-surface sheet.  Similar conclusions were drawn from STM/STS measurements \cite{WatashigeSTMPRX2015},  while the absence of a sizable residual $\kappa$/T in heat-transport measurements \cite{TailleferFeSe2016} allows us to rule out symmetry-imposed nodes only. Since the crystal structure has been reduced to C$_{2}$-symmetry by the transition at T$_{s}$,  the superconducting gap structure cannot be classified in a pure $s$-or $d$-wave symmetry. However, a real $s+d$ admixture, {\it i.e.} $\Delta({\bf k})=\Delta_{s}+\Delta_{d}\cdot \cos(2\theta$), is likely realized in FeSe due to the close proximity of $s\pm$ and $d$-wave instabilities \cite{Fernandes2013PRL,FanPRB2013,KretzschmarPRL2013}. Indeed, recent theoretical calculations \cite{KangPRL2014} showed that the ratio $\Delta_{s}/\Delta_{d}$ can be effectively tuned by changing the $d_{xz}/d_{yz}$ splitting so that the existence of nodes is related to the amount of strain in the crystal. 
 Evidence for a very small gap have been presented in recent thermal conductivity  \cite{TailleferFeSe2016} and penetration depth measurements \cite{MLiarXiv2016}. Heat capacity measurements to even lower temperatures (100 - 200 mK) would be needed to observe such a small gap.

Figure \ref{Fig3} demonstrates that , in contrast to previous reports \cite{MahmoudPRB2015}, S-substitution leads to drastic changes of the low-energy excitations observed at low temperature. In particular, we find evidence in our crystals for a significant residual C$_{e}$/T term at low temperatures for $x$ = 0.15, in analogy to what has been found in Co-doped Ba122 \cite{FredEurophyL2010} and which has been interpreted in terms of strong pair-breaking effects \cite{FredarXiv2016}.  The presently observed residual ungapped density of states may also be a sign of pair-breaking due to interband scattering in a $s\pm$ scenario \cite{PhysRevB.55.15146}. Since FeSe$_{1-x}$S$_{x}$ is orthorhombic, the average value of the energy gap over the Fermi surface $<\Delta({\bf k})>_{FS}$ does not vanish, and therefore the sign change occurs between hole and electron pockets as expected in a $s\pm$ state.  In this scenario, T$_c$ would however be expected to decrease for significant interband scattering instead of the experimentally observed increase. This discrepancy possibly may be due to the increase in $\gamma_n$ with S substitution, which would be expected to increase T$_c$.

\paragraph{Pressure dependence of T$_c$}

The discontinuities, $\Delta C$ and $\Delta \beta$ at T$_c$ can be used to calculate the hydrostatic pressure dependence of $T_c$ using the Ehrenfest relation
\begin{equation}
\label{eq3}
\frac{dT_c}{dp_i}=V_m\frac{\Delta \beta}{\Delta C/T_c}
\end{equation}  
Here $V_m$ is the molar volume, which in our case is 23.34 cm$^3$ mol$^{-1}$. By using the data from Fig. \ref{Fig1}(b) and Fig. \ref{Fig2}(c), we get $dT_c/dp$ = 6.4 and 4.0 K/GPa for pure and 15$\%$ doped FeSe respectively.  The decrease in $dT_c/dp$ with S substitution is consistent with the decrease of $dT_c/dp$ observed for increasing pressure in actual pressure measurements \cite{TerashimaPRB2016,TerashimaJPSJ2015,UdharaPRB2016}, demonstrating that S-substitution acts as a chemical pressure.

\section{Conclusions}

In conclusion, using thermal expansion and heat capacity measurements, we have investigated the thermodynamic properties of pure and S-substituted FeSe single crystals.  Surprisingly we find that, in strong contrast to other Fe-based systems, superconductivity in FeSe$_{1-x}$S$_x$ favors the distorted nematic state, which should provide further clues about the origin of the spin-versus-orbital debate in these materials.  Our heat capacity data provide evidence for a near nodal superconducting gap structure for pure FeSe.  With S substitution, the nodal behavior disappears and a large residual low-temperature density of states emerges, which may be understood in terms of intra- and interband scattering in sign changing $s\pm$  gap structure.

\begin{acknowledgement}
We acknowledge useful discussions with Amalia Coldea, Matt Watson, Peter J. Hirschfeld and Ilya Vekhter.
\end{acknowledgement}

%
\bibliographystyle{pss}
\bibliography{SPPFeSe}



\end{document}